\documentclass[aip,apl,reprint,showpacs,showkeys,twocolumn,superscriptaddress]{revtex4}
\usepackage[pdftex]{graphicx}
\usepackage{ulem}
\usepackage{amssymb}
\usepackage{bm}
\usepackage{color,multirow}
\preprint{KIST-XXX}

\newcommand\mm{{Fe/MgO/Fe/Co\ }} 
\begin{document}
\setcounter{page}{1}
\title[Magnetic and electronic structure of \mm multilayer stack]{Magnetic and Electronic Structure study of \mm Multilayer Stack Deposited by E-Beam Evaporation}

\author{Jitendra Pal \surname{Singh}}
\affiliation{Inter-University Accelerator Centre, Aruna Asaf Ali Marg, New Delhi 110 067, India}

\author{Sanjeev \surname{Gautam}} \email{sgautam71@kist.re.kr (S. Gautam)} 
\affiliation{Advanced Analysis Center, Korea Institute of Science and Technology (KIST), Seoul 136-791, Republic of Korea}

\author{Braj Bhusan \surname{Singh}} \author{M. \surname{Raju}} \author{S. \surname{Chaudhary}}
\affiliation{Department of Physics, Indian Institute of Technology, New Delhi 110 016, India}

\author{D. \surname{Kabiraj}} \author{D. \surname{Kanjilal}}
\affiliation{Inter-University Accelerator Centre, Aruna Asaf Ali Marg, New Delhi 110 067, India}

\author{Jenn-Min \surname{Lee}} \author{Jin-Ming \surname{Chen}}
\affiliation{National Synchrotron Radiation Research Center, Hsinchu 30 076, Taiwan}

\author{K. \surname{Asokan}}
\affiliation{Inter-University Accelerator Centre, Aruna Asaf Ali Marg, New Delhi 110 067, India}

\author{Keun Hwa \surname{Chae}} \email{khchae@kist.re.kr(K.H. Chae)} \thanks{FAX:+82-54-279-1599}
\affiliation{Advanced Analysis Center, Korea Institute of Science and Technology (KIST), Seoul 136-791, Republic of Korea}

\date{\today}
\begin{abstract}
Present work investigates the magnetic and electronic structure of MgO/Fe/MgO/Fe/Co/Au multilayer stack grown on Si(100) substrates by electron beam evaporation method. X-ray diffraction study depicts polycrystalline nature of the multilayers. Results obtained from vibrating sample magnetometry (VSM) and near-edge X-ray absorption fine structure spectra (NEXAFS) at Fe \& Co L- and Mg \& O K-edges are applied to understand the magnetic and electronic properties of this stack and its interface properties. While the spectral features of Fe L-edge spectrum recorded by surface sensitive total electron yield (TEY) mode shows the formation of  FeO$_x$ at the Fe/MgO interface, the bulk sensitive total fluorescence yield (TFY), shows Fe in metallic nature. Co L-edge spectrum reveals the presence of metallic nature of cobalt in both TEY and TFY modes. Above results are well correlated with X-ray reflectometry.
\end{abstract}


\maketitle

Devices of multilayer structures are important in the development of spintronic devices based on giant magnetic resistance (GMR) and tunneling magnetorsistance (TMR), thermoelectric applications\cite{rr1,r2,r4,rr5,r6,r7,rr8,r9,r10}. TMR based devices, {known as magnetic tunnel junction(MTJ),} require MgO as a barrier layer and find applications in magnetic random access memories (MRAM) \cite{r4,r5}. It has been shown that properties of these devices are sharply influenced by the interface composition and roughness of the multilayer stack. Till date 604\% TMR has been reported \cite{r9} which is less than the theoretically predicated value of TMR ($\sim$ 1000\%) \cite{r10,r11} and this discrepancy has been attributed to the interface properties of the heterostructures \cite{r11,rr11}. Hence, there is a need to investigate the growth and interfaces properties of Fe/MgO/Fe like structures in order to get a correlation among the interface and its bulk properties.  {Spectroscopic techniques like near edge X-ray absorption fine structure (NEXAFS) has emerged as a powerful tool for determination of local chemical structure. The technique is also sensitive to small strain induced variations in bond lengths \cite{ref29} and efficient to detect atomic interdiffusion processes at the interfaces \cite{ref31}. These aspects have been investigated by several authors in CoFeB/MgO/CoFeB structures \cite{r12,r13,r14,r15}. Conclusion obtained from these various study lead to utilization of several methods which include insertion of pinning layers and annealing the whole structure in high vacuum for improving interface roughness and crystallinity of this multilayer stack} \cite{r16,r17,r18}. Besides the post deposition treatment, deposition methods also play important role in determining the properties of the multilayer stacks. Although methods like molecular beam epitaxy (MBE) and rf-sputtering are commonly employed, the most cost effective e-beam evaporation method may be an alternative tool for deposition of this type of multilayers \cite{r19}. The present work reports the deposition of Fe/MgO/Fe/Co/Au multilayer stack by e-beam evaporation and the local electronic structure by using NEXAFS techniques. A layer of Co was deposited above Fe layer to check the magnetic properties. Interface properties are discussed and correlated with magnetic and electronic properties.
\section{Experimental Details}
\subsection{Deposition of multilayer structures}
Multilayer stack of MgO/Fe/MgO/Fe/Co/Au was deposited on Si(100) substrates by e-beam evaporation method with base pressure better than $2\times10^{-8}$ Torr. Si substrates were cleaned in trichloroethylene, acetone, isopropyl alcohol and de-ionized water in order to remove chemical impurity and surface contamination. Si substrate was also dipped into hydrogen fluoride (HF) for 2-3 minute to remove any native SiO$_2$.  For the deposition of MgO layers, MgO powder (purity $99.999$\%, Alfa-Aesar) was pressed into pellet form and evaporated. Fe, Co and Au targets were used for metal layer deposition. All these targets were kept inside the chamber, so that deposition of all the layers could be done without breaking vacuum. This is essential to avoid any contaminations of interfaces. All the layers were deposited with deposition rate of $0.1$ nm/s with online monitoring of thickness using quartz crystal monitor. First, MgO buffer layer was deposited on Si(100) substrate in order to prevent silicide formation at Si/Fe interface due to diffusion. MgO layer also works as a good buffer layer to grow epitaxial Fe layer\cite{r10}. The substrate temperature was kept at $300~ ^\circ$C during deposition. On top of MgO buffer layer, Fe thin film was deposited at 180 $^\circ$C. Subsequently, MgO barrier layer was deposited at the same growth temperature. The upper Fe layer was deposited at temperature of $200$ $^\circ$C. Further Co layer was deposited at $200~ ^\circ$C. This multilayer structure was annealed at $315$ $^\circ$C for $1.5$ hr in ultra high vacuum to improve crystallinity and make interfaces sharp. After annealing a capping layer of Au was deposited at room temperature in order to prevent Co from oxidation.
\begin{figure}[tbh!]\centering
  \includegraphics[width=8cm]{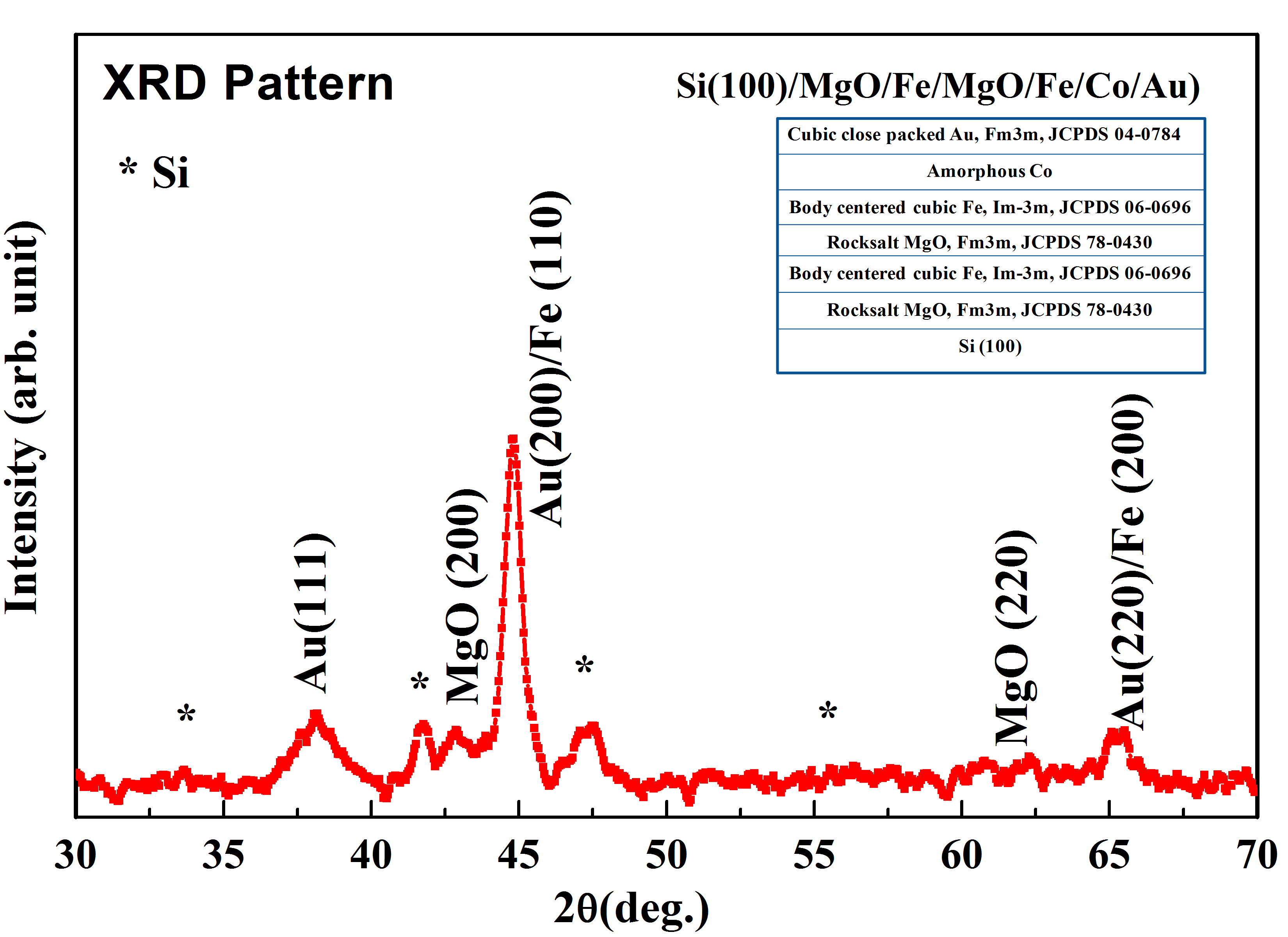}
  \caption{X-ray diffraction pattern of Fe/MgO/Fe/Co/Au multilayer. Inset shows layer sequence along with crystal structure of each layer (not to scale). }\label{fig1}
\end{figure}
\subsection{Characterization Techniques}
Rutherford backscattering spectroscopic (RBS) study for these multilayer structure was performed on RBS spectrometer (Inter University Accelerator Centre (IUAC), New Delhi).  {It is observed that all the layers exhibit good stoichimety and the detailed study has been published elsewhere \cite{r20}.} X-ray diffraction study on multilayer structure was carried out on a Philips X-ray diffractometer {with step time of 2.5 s and step size of 0.05$^\circ$ was used for recording the XRD pattern.} Magnetic studies on the multilayers were carried out using a vibrating sample magnetometer (VSM) at National Physical Laboratory, New Delhi. Further, electronic structure was investigated by near-edge X-ray absorption fine structure (NEXAFS) at the high energy spherical grating monochromator (HSGM) BL20A1 beamline in the National Synchrotron Radiation Research Center (NSRRC) in Taiwan. All measurements were processed in an ultra high vacuum (UHV) chamber ($\sim 10^{-9}$ Torr) at $300$ K, in (a) total electron yield (TEY, surface sensitive), measured by monitoring the total sample photocurrent,  and (b) total fluorescence yield (TFY, bulk sensitive), measured with a negatively biased microchannel plate(MCP), simultaneously. Measurements in TEY and TFY modes probe the electronic structure of materials with a probing depth $\sim$ 4 nm and several hundred nanometers, respectively \cite{stohr96}.
Around the edge energy in NEXAFS spectrum, data points with an energy spacing of $0.02$ eV were recorded, using a $1$ sec. collection interval per point. The incoming radiation flux (I$_0$) was monitored by the total photocurrent produced in a highly transmissive Au-mesh inserted into the beam. The overall photon resolution around O K-edge was $\sim 0.06$ eV using $15\times15\mu$m slits. After a constant background substraction, all spectra were normalized to the post-edge step height using Athena 0.0.061 \cite{athena}. X-ray reflectometry (XRR) has also been carried out for these structures and the results obtained were simulated by using GenX 2.0b reflectivity software \cite{genx,rr25}.
\begin{figure}[tbh!]\centering
  \includegraphics[width=8cm]{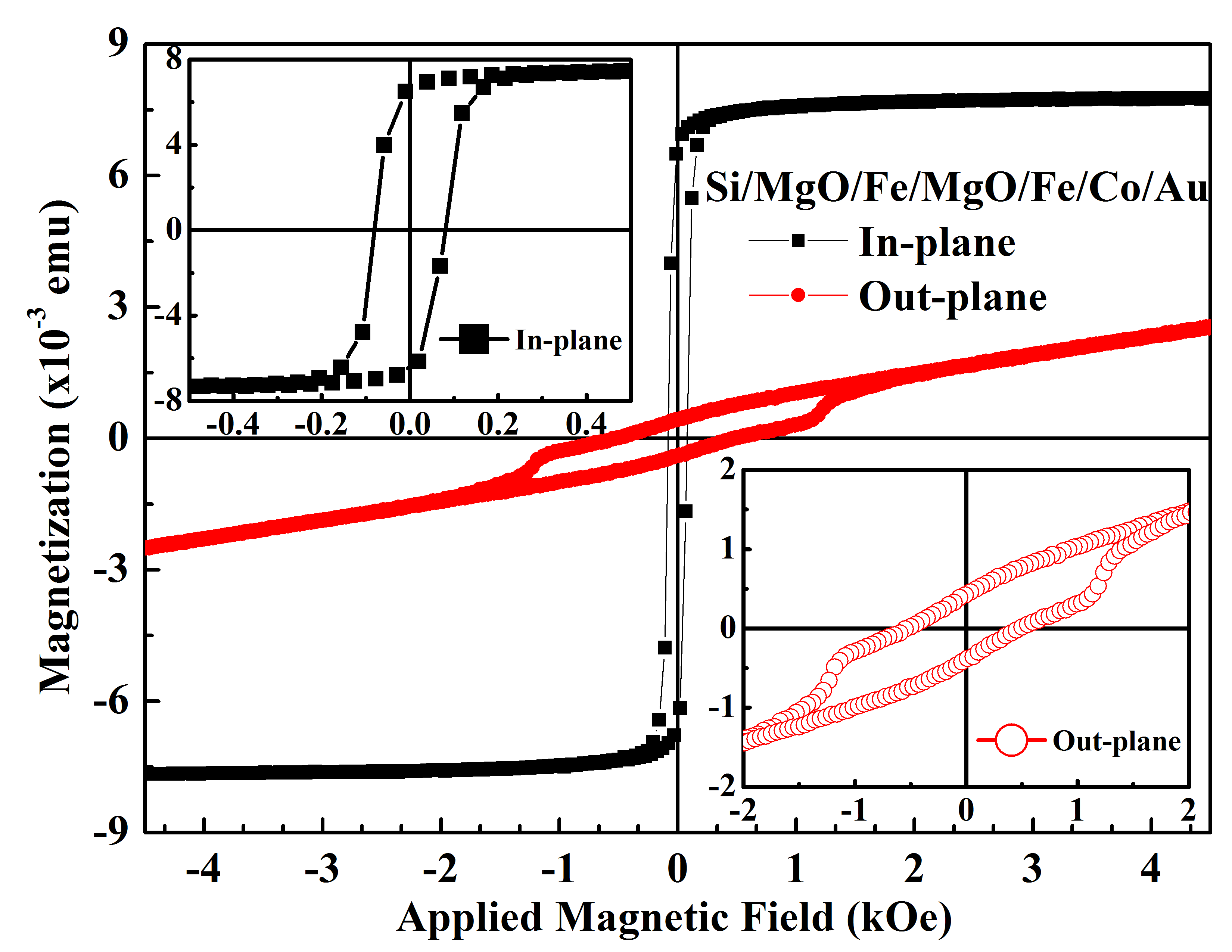}\\[-0.1cm]
  \caption{Hysteresis curves of Fe/MgO/Fe/Co/Au multilayer in parallel (in) and perpendicular (out) direction. In-set clearly shows the change in hysteresis shape for in-plane and out-of-place configuration. }\label{fig2}
\end{figure}
\section{Results and Discussion}
\subsection{Structural and Magnetic study}
Figure \ref{fig1} shows the XRD pattern of Fe/MgO/Fe/Co multilayer. The peaks in the XRD pattern has been identified to come from body centered Fe [JCPDS-06-0696], rocksalt MgO [JCPDS-78-0430], and cubic close packed Au [JCPDS 04-0784]. No peak is observed corresponding to the phase of Co. Hence, it may be contemplated that Co layer is amorphous in nature. For better visualization of structure of various layer and their sequence, a schematic figure has been shown in Fig \ref{fig1}:inset.
In Table \ref{tab1} we have shown the thickness of various layers as simulated by RBS. Figure \ref{fig2} shows the magnetic hysteresis curves of the multilayer stack while applying the magnetic field parallel and perpendicular to the film surface. Both these hysteresis are almost symmetric. Parallel hysteresis is very much similar to the pure bcc Fe \cite{r22}.  It saturates only at the value of $1500$ Oe and has the value of saturation magnetic moment(m$_s$) $\sim 7.4\times10^{-3}$ emu. The loop is also closed at both ends.  The coercivity(H$_c$) and remanence(m$_r$) are $80$ Oe and $6.5\times10^{-3}$ emu, respectively. The hysteresis curves measured in perpendicular does not saturate upto $5$ kOe and have almost paramagnetic-like behavior after $2$ kOe. The coercivity and remanence of this stack are $505$ Oe and $\sim 0.42\times10^{-3}$ emu, respectively. This hysteresis exhibits two step behaviour, which is may be due to presence of insulating layer (MgO) between two ferromagnetic (FM) layers. Absence of sharp steps in this hysteresis may be due to polycrystalline nature and oxidation of Fe. This multilayer stack exhibits the in-plane anisotropy with squareness ratio m$_r$/m$_s \sim$ 1.
\begin{figure}[tbh!]\centering
  \includegraphics[width=8cm]{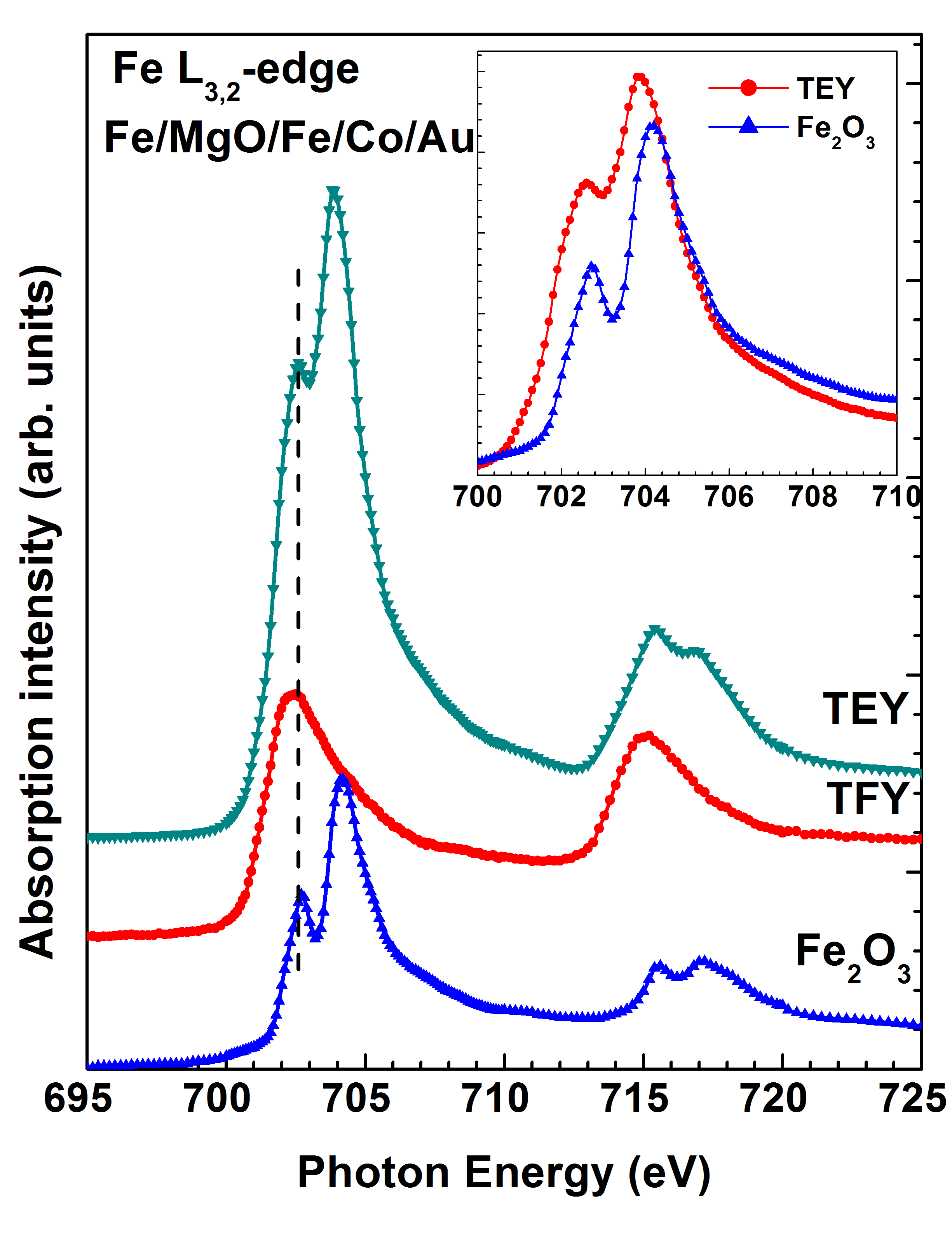}\\[-0.1cm]
  \caption{Normalized NEXAFS spectra at Fe L$_{3,2}$-edge for Fe/MgO/Fe/Co/Au multilayer structure plotted along with Fe$_2$O$_3$(Fe$^{3+}$) as reference. Surface sensitive TEY mode shows the close resemblance with Fe$_2$O$_3$ and bulk sensitive TFY mode shows the presence of Fe-metal in the depth of Fe layers.}\label{fig3a}
\end{figure}
\begin{figure}[tbh!]\centering
  \includegraphics[width=8cm]{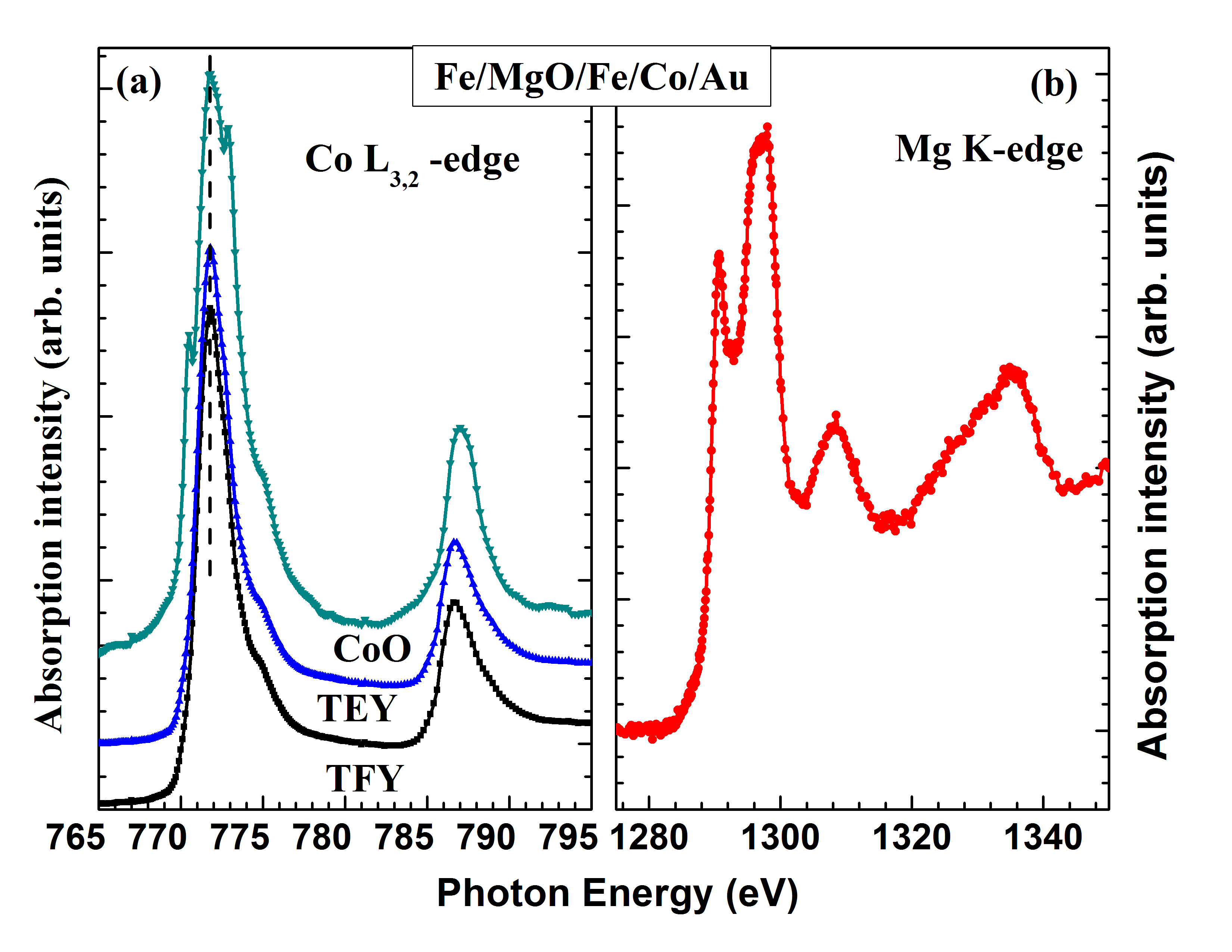}\\[-0.1cm]
  \caption{Normalized NEXAFS spectra at (a) Co L$_{3,2}$-edge for Fe/MgO/Fe/Co/Au multilayer (in TEY and TFY modes) along with CoO reference spectrum collected simultaneously. (b) Mg K-edge for Fe/MgO/Fe/Co/Au multilayer.}\label{fig3bd}
\end{figure}
\subsection{Near Edge X-ray Absorption Fine Structure (NEXAFS) study}
Figure \ref{fig3a} shows the Fe L-edge NEXAFS spectra for the multilayer system measured in TEY and TFY mode along with Fe$_2$O$_3$ as a reference.
{NEXAFS spectra recorded for Fe L-edge in TEY mode show the peaks at 703, 705, 716 and 718 eV.} The Fe L-edge spectrum arises due to Fe 2$p$ core level, which in effect of spin-orbit coupling, give rise to degenerate state 2$p_{3/2}$ and 2$p_{1/2}$ showing multiplets centered on $704$ and $717$ eV. These octahedral crystal fields lifts the degeneracy of the 2$p_{3/2}$ and 2$p_{1/2}$ levels so that two levels with t$_{2g}$ and e$_g$ symmetry are created, as indicated by the two structures at about $703$ and $705$ eV and at $716$ and $718$ eV. These structures are indicative of Fe$^{3+}$ oxidation state and generally observed in Fe-based oxide systems \cite{r23}. Since, TEY mode is surface sensitive, hence it may be contemplated that MgO/Fe interfaces in multilayers have oxidized Fe at interface. The oxidation of upper Fe layer cannot expect in the present case because this layer is capped by Co and Au layer on top. The bulk sensitive TFY mode Fe L$_{3,2}$-edge spectrum exhibits peaks at $708$ and $714$ eV and indicates that Fe is not bonded with oxygen. This shows the metallic nature of Fe. Mg K-edge spectrum exhibits peaks around the position $1284$, $1300$, $1308$ and $1332$ eV (Fig. \ref{fig3bd}(b) that are associated primarily with Mg 1$s$ to 3$p$ states \cite{r24} and similar to MgO spectrum.
\begin{figure}[tbh!]\centering
  \includegraphics[width=8cm]{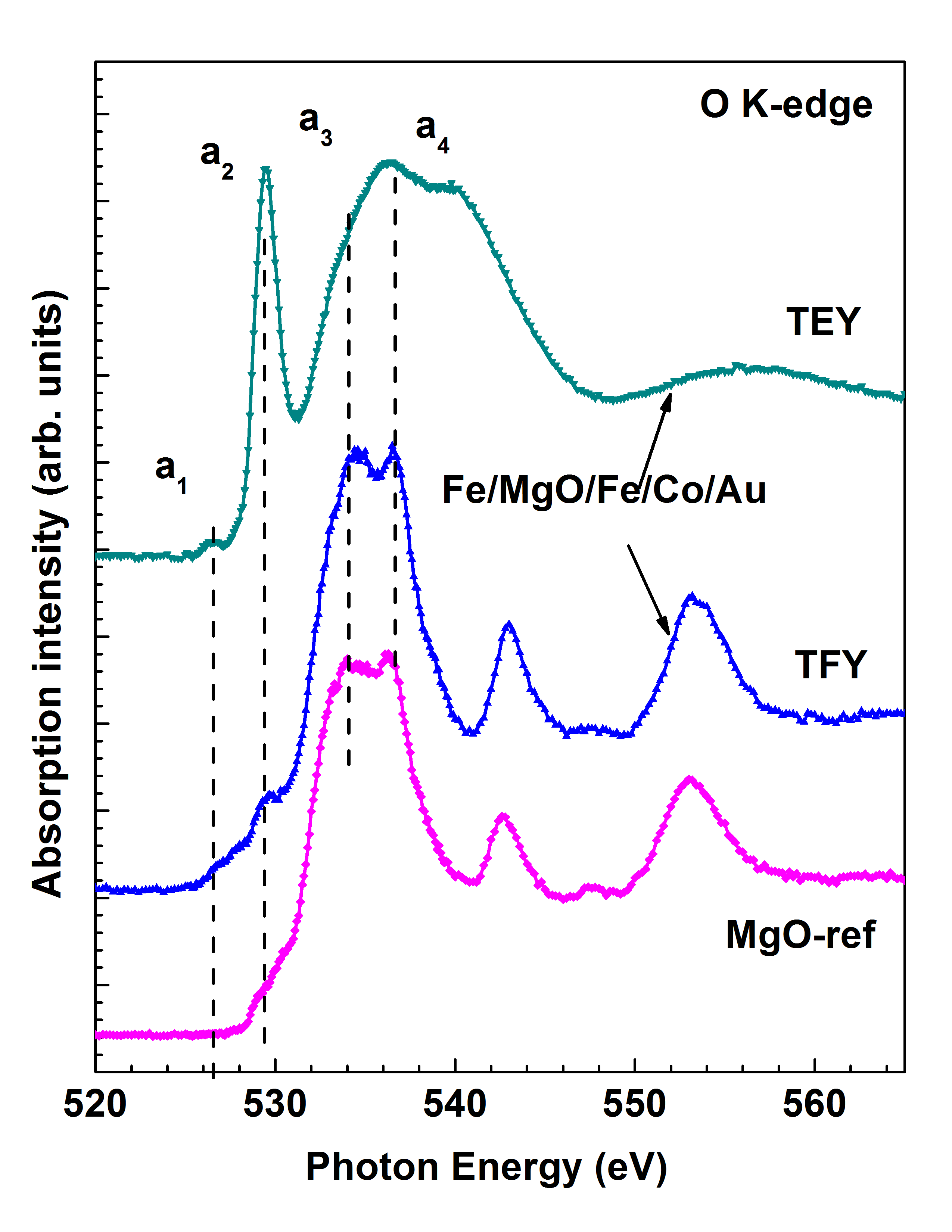}\\[-0.1cm]
  \caption{Normalized NEXAFS spectra at O K-edge for Fe/MgO/Fe/Co/Au multilayer in TEY and TFY mode plotted along with he MgO reference spectrum collected simultaneously.}\label{fig3c}
\end{figure}

Figure \ref{fig3c} shows O K-edge NEXAFS spectra in TEY and TFY mode of the multilayer along with MgO for comparison. Out of these, TFY spectrum consists of pre-edge structure around $\sim 529.83$ eV  and these structures generally due to the excitation to the localized bound state and consistent with the previous reports \cite{r25}. The observed spectral features in this spectrum, around $534$, $536$, $542$ and $552.9$ eV are very much similar to the spectrum of MgO thin films reported by Luchs {\it et\ al.\ } \cite{r26}. A visual inspection shows large difference in the TEY and TFY spectra of the samples. {In TEY spectrum,} we observe peaks around $\sim$ $527$, $530$, $534$, $538$, $545$ eV. These features are very much similar to ferrites \cite{r27,r28}, where the spectra originate from transitions into unoccupied states with O 2$p$ character hybridized with metal ions. This is in accordance with the formation of Fe-O$_x$ at the Fe/MgO and MgO/Fe interfaces. Fig. \ref{fig3bd}(a) shows the Co L-edge spectrum of Fe/MgO/Fe/Co/Au structure with the peaks occurring at $772$ and $776$ eV. Spectral feature is very  much analogues to the metallic Co \cite{r29}. The spectra have the same intensity in both the TEY and TFY mode. This indicate that Co atom was not bonded with Fe atom consistent with RBS study \cite{r20}, however the possibility of diffusion of Co metal into Fe cannot be neglected.  {The results obtained from NEXAFS spectra confirms the assumption of possible oxidation of Fe layers at the interface.}
\begin{table}\begin{center}\scriptsize
\caption{Thickness (t) estimated from Rutherford Backscattering Spectroscopy (RBS) simulation and X-ray reflectivity (XRR) simulation parameters for density ($\rho$), thickness (t) and roughness ($\sigma$) of various layers in Fe/MgO/Fe/Co multilayer structure}
\vskip 0.5cm \begin{tabular}{|c|c|c|c|c|} \hline
\multirow{2}*{Layers} & RBS & \multicolumn{3}{c|}{XRR}  \\ \cline{2-5}
       &t & $\rho$ & t$\pm$0.1 & $\sigma\pm$0.01 \\
       &(nm) & (gm/cm$^3$) & (nm) & (nm)  \\ \hline
Si     & -- & 2.33  & - & 0.17 \\ \hline
SiO$_x$& -- & 1.96  & 1.9& 0.97 \\ \hline
Mg$_2$Si$_x$& -- & 1.09  & 1.1& 0.60 \\ \hline
MgO& 85.0 & 3.55  & 50.1& 0.41 \\ \hline
FeO$_x$& -- & 5.08  & 1.1& 1.28 \\ \hline
Fe & 34.0 & 7.21  & 48.8& 0.48 \\ \hline
FeO$_x$& -- & 3.49  & 2.2& 1.99 \\ \hline
MgO & 24.0 & 3.45  & 26.7& 0.46 \\ \hline
FeO$_x$& -- & 5.04  & 1.0& 1.46 \\ \hline
Fe & 12.0 & 7.27  & 34.8 & 0.72 \\ \hline
Fe-Co & 18.0 & 7.45  & 59.1& 1.26 \\ \hline
Co & 43.0 & 7.18  & 59.1& 1.26 \\ \hline
Au & 4.2 & 10.61  & 0.8 & 2.88 \\ \hline
\end{tabular}\label{tab1}\end{center}\end{table}
\subsection{X-Ray Reflectivity(XRR) study}
\begin{figure}[tbh!]\centering
\includegraphics[width=8cm]{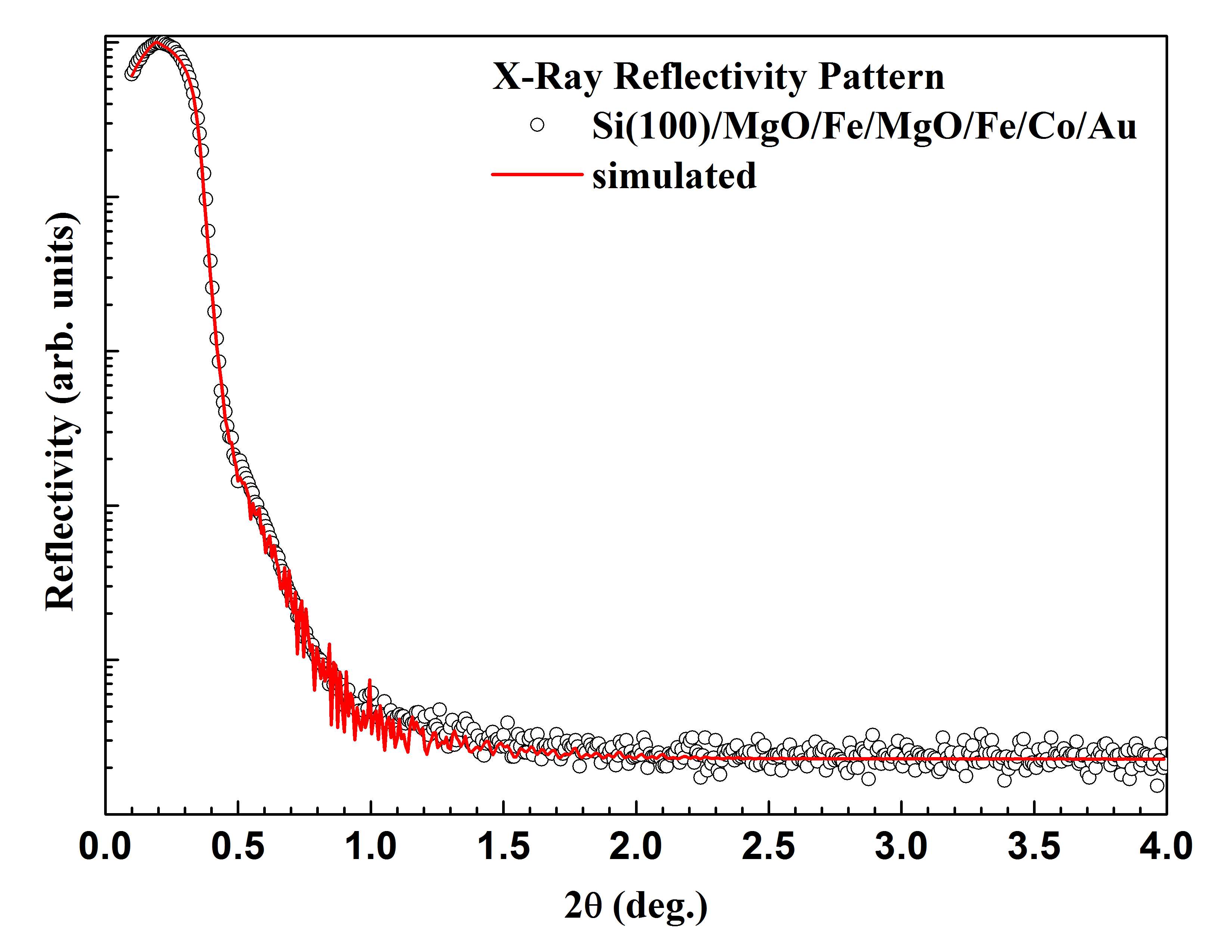}\\[-0.1cm]
\caption{Experimental and Simulated XRR curves for Fe/MgO/Fe/Co/Au multilayer.}\label{fig4}
\end{figure}

Above results obtained from VSM and NEXAFS analysis provide strong evidence for oxidation at Fe/MgO/Fe interface and further characterization of thickness and interface was carried out by using XRR. Figure \ref{fig4} shows XRR pattern for multilayer and the fitted curves. Parameters used to simulate the experimental data are given in Table \ref{tab1}. Data from the Table shows that the thickness of layers  {is different from that are obtained from RBS study.} Besides this formation of Fe-Co alloy also expected at Fe/Co interface. This corroborates results obtained from the RBS analysis discussed elsewhere \cite{r20}. The barrier layer of multilayer is $\sim 22$ nm, which is very large compared to conventional MTJ stack. Due to this large thickness, de-coupling between two FM layers is expected and different coercivities of upper and lower FM layers should be clearly observed but
 {this effect could not be observed in the hysteresis curve of multilayer.} It is expected that coercivity is also affected by the polycrystalline nature of various layers and oxidation at interfaces inhibiting the presence of magnetic switching between these two layers.   {Coercivity of any magnetic thin film depends on microstructure of film, thickness and also the orientation of film growth \cite{r29,r30,r31}. A detailed investigation on Fe film shows that due to presence of magnetocrystalline anisotropy, coercivity is different for different plane \cite{r32}.   Hence, it may be contemplated that due to same nature of growth both the Fe film have almost same value of coercivity.}
\section{Conclusions}
Fe/MgO/Fe/Co/Au multilayer stacks were deposited by e-beam evaporation method in ultra high vacuum. Magnetization of this stack is similar to the pure bcc Fe, even though a thick MgO barrier is present between the ferromagnetic layers. NEXAFS study performed on this stack shows the oxidization of Fe/MgO interface, which is also confirmed by XRR.  Metallic nature of Fe (bulk) and Co layers were observed by using NEXAFS study.

\begin{acknowledgments}
JPS is thankful to Prof. J. S. Moodera, MIT USA for fruitful discussion during the fabrication of multilayer stack. Authors are also thankful to Mr. S.R. Abhilash for providing experimental support during the e-beam deposition of stack. Synchrotron experiments at NSRRC, Taiwan are supported by the KIST (Grant No. 2V02631) and the Korean Synchrotron User Association (KOSUA).
\end{acknowledgments}

\bibliographystyle{apsrev}
\bibliography{FeCoMM2}

\end{document}